\documentclass[twocolumn]{aastex701}
\usepackage{times}
\usepackage{natbib}
\usepackage{hyperref}
\usepackage{graphicx}
\usepackage{dcolumn}
\usepackage{xcolor}
\usepackage{amsmath}
\usepackage{amssymb}
\usepackage{bm}

\begin{document}

\title[Halo-driven Origin and Evolution of SMBHs]{Halo-driven Origin and Evolution of Overmassive Black Holes and Little Red Dots
}

\author[orcid=0009-0001-3469-9812,gname='Ritik',sname='Sharma']{Ritik Sharma}
\affiliation{Department of Physics, Indian Institute of Technology (IIT) Bhilai, Chhattisgarh, 491002, India}
\email{ritiks@iitbhilai.ac.in}  

\author[orcid=0000-0001-9927-5255,gname='Mahavir', sname='Sharma']{Mahavir Sharma*} 
\affiliation{Department of Physics, Indian Institute of Technology (IIT) Bhilai, Chhattisgarh, 491002, India}
\email[show]{mahavir@iitbhilai.ac.in}

\begin{abstract}
We present a theoretical model in which the recently detected overmassive black holes (OBHs), and possibly Little Red Dots (LRDs), arise during a halo-driven transient phase preceding the established coevolution of supermassive black holes (SMBHs) and their host galaxies. In this model, halo gravity drives an early phase of rapid black hole growth, leading to systems in high-redshift haloes that lie above the local scaling relations. As the halo evolves, a transition in halo thermodynamics leads to the onset of a hot mode that suppresses accretion, reducing the black hole growth rate and driving the system toward the local black hole mass–stellar mass relation. LRDs may represent an observational manifestation of the rapid, halo-driven growth phase, while OBHs trace its direct mass signature. Our model thus provides a unified framework in which these systems form and evolve toward the regulated coevolution observed in the local Universe.
\end{abstract}

\section{Introduction} 
\cite{Schmidt1963} reported a new class of objects showing star-like characteristics but unusually high luminosities. These compact sources were extremely distant and among the most luminous objects in the Universe, and were named `quasars' or Quasi Stellar Objects (QSOs). The properties of QSOs posed a theoretical challenge as no known stellar process could account for such enormous energy output from a compact region. The initial theoretical models suggested that QSOs were supermassive stars \citep{HoyleFowler1963}. However, later these objects were explained by models in which the nuclear part of a galaxy is powered by a compact central engine, now understood to be a supermassive black hole (SMBH) \citep{Salpeter1964,Zeldovich1964, Linden1969, Urry1995}. The direct detections of supermassive black holes in galactic centres established that nearly all massive galaxies host central SMBHs \citep{Kormendy95}.

The SMBHs at galaxy centres typically have masses in the range $10^6$ - $10^9$ $\rm M_\odot$ \citep{Mortlock2011, Kormendy13, Banados16, Inayoshi2020, Fan2023}, which is extremely high compared to stars and black holes in binary systems and can be as high as $\sim 0.1 \%- 1 \%$ of the galactic components such as the bulge or disc \citep{Kormendy13}. \cite{Magorrian1998} reported a correlation between black hole mass and bulge mass. Later studies revealed an even tighter correlation between the black hole mass and the bulge velocity dispersion, known as the $M_{\rm bh}$–$\sigma$  relation \citep{Tremaine02}. The galaxy and its components such as the bulge, disc and SMBH, they all form from baryons cooling and sinking in the potential wells of the host dark matter halo. Indeed, \cite{Ferrarese02} reported a correlation between SMBHs and their host haloes, though it has been hotly debated \citep[e.g.][]{Kormendy_2011,Kormendy13}.

In the last decade, direct imaging with facilities such as the Hubble Deep Field (HDF) \citep{HubbleDeep2006}, Sloan digital Sky Survey (SDSS) \citep{York2000}, Subaru \citep{Subaru2021}, Atacama Large Millimeter/submillimeter Array (ALMA) and the James Webb Space Telescope (JWST) \citep{JWST2006} has found a considerable population of SMBHs at $z>6$ \citep{Inayoshi2020, Fan2023, Larson2023, Maiolino2024a, Maiolino2024b, Yue2024}. The existence of SMBHs that early in cosmic history is a theoretical challenge, and the current models rely on super-Eddington accretion \citep{Bromm03, Johnson07, Madau2014}, the formation of heavy seeds through runaway stellar collisions in dense star clusters \citep{Portegies2004, Devecchi2009} or the direct collapse of gas clouds to create heavy seeds \citep{Bromm03, Begelman2006, Lodato2006}  (see \citet{Inayoshi2020, Greene2020} for detailed reviews). Moreover, the high-redshift black holes seem to be in tension with the local correlations between black holes and their host galaxies \citep{Harikane2023, Yue2024, Li2025, Juodzbalis2026}. 

Cosmological simulations have explored the coevolution of SMBHs with their host haloes and galaxies \citep{Booth_2010, Volonteri_2011, Bogdan_2015}. The halo likely plays an important role in growth of its central black hole \citep{Park2017, Ciotti18}, and it can drive a rapid growth of the central black hole to SMBH scales, that can explain the recent JWST observations \citep{Sharma_2024}. However, once a seed black hole transitions to an SMBH rapidly at high redshift, the growth rate must decrease; otherwise, the SMBH mass would become unrealistic. The decrease in growth rate is also desirable for explaining local coevolutionary relations, as the observed growth in the present-day Universe appears to be self-regulated. Several possible mechanisms have been discussed in the past to cause a decrease in the growth rate, for example, the quenching in growth rate due to radiative feedback from accreting gas \citep{Silk98, King03} and the onset of star formation and resulting starvation of the central black hole \citep{Hopkins2011, Angles17}.

The recent discoveries of overmassive black holes (OBHs) \citep{Harikane2023, Stone2024, Yue2024}, defined as systems lying significantly above the local BH-stellar mass relation have challenged the co-evolution paradigm. Further, JWST has recently discovered a new class of compact, red sources known as Little Red Dots (LRDs) in the high-redshift Universe, and their origin is under debate. The recent observations support the view that LRDs are obscured Active Galactic Nuclei (AGN)  with rapidly growing massive black holes \citep{Labbe2025, Pacucci2025, Akins2025, Hviding2026, Rusakov2026, Pang2026}. In many cases, the inferred black hole masses place them in the category of OBHs \citep{Chang2025, Pang2026}. These high-redshift populations likely benefit from an unrestricted gas supply in dense environments within early halos \citep{Pacucci2025, Rusakov2026, Pang2026}, which aligns with the view that parent halos initially prioritise the growth of the central black hole \citep{Sharma_2024}. However, this rapid growth spurt must eventually be curtailed to avoid a catastrophic growth sprint, and to reconcile high-redshift OBHs with the regulated coevolution seen at lower redshifts. 

In this paper, we explore a halo-driven mechanism to explain the early growth of SMBHs and the emergence of OBHs. 
In general, the formation of SMBHs is intertwined  with galaxy formation, through a complex interplay of gas infall, cooling, angular momentum, star formation and feedback, and is usually addressed with full cosmological simulations \citep[e.g.][]{Sijacki2015, Schaye15, Dave2019, Ni2022}. Our objective here is to isolate the fundamental physical aspects of the underlying hydrodynamics in a combined black hole and halo gravitational system. In doing so, we uncover a halo-driven, macro-scale regulatory mechanism that provides a unified evolutionary picture, beginning with the transition to halo-gravity-driven rapid growth to explain high-redshift SMBHs, followed by halo-thermodynamics-driven regulation as the host halo surpasses a critical mass, resulting in the transition from `cold mode' accretion to a pressure-supported `hot mode' \citep[e.g.][]{Dekel2006} that steers these systems toward local scaling relations. We specifically test how 
this `spurt-and-quench' framework accommodates the recent JWST observations of OBHs and LRDs.

In Section~\ref{sec2phase}, we describe the halo-driven transition to rapid black hole growth and compare our results with JWST observations.
In Section~\ref{sec_suprresion}, we introduce the halo-thermodynamics-driven suppression. In Section~\ref{sec_result}, we analyse the resulting turnover in the $M_{\rm bh}\hbox{--}M_\star$ and $M_{\rm bh}\hbox{--}M_{\rm h}$ planes. We summarise our findings and discuss the implications for galaxy-SMBH coevolution in Section~\ref{sec_discussionC3}.

\section{Halo-Gravity driven rapid growth}
\label{sec2phase}
We model the early growth of SMBHs using the steady-state hydrodynamic accretion into a combined black hole and halo  potential \citep{Sharma_2024}.
The steady, spherically-symmetric flow is governed by,
\begin{equation}
 \frac{d}{dr} (\rho v r^2) = 0 \label{massC3}
\end{equation}
\begin{equation}
    v\frac{dv}{dr} = -\frac{1}{\rho}\frac{dp}{dr} - \frac{d\phi}{dr} \ . \label{momentC3} 
\end{equation}
Here $\rho$ is the density, $v$ the velocity, $p$ the pressure and $\phi$ is the combined gravitational potential due to a dark matter halo and a central black hole. For the dark matter, we use a Navarro-Frenk-White (NFW) dark matter profile \citep{NFW1997}. By using the Mach number, $\mathcal{M}=v/c_s$, and sound speed $c_s = \sqrt{dp/d\rho}$, Eq.~\ref{momentC3} transforms to,   
\begin{equation}
    \frac{\mathcal{M}^2-1}{2\mathcal{M}^2}\frac{d\mathcal{M}^2}{dr} + \frac{\mathcal{M}^2-1}{2c_s^2}\frac{dc_s^2}{dr} = \frac{2}{r} - \frac{1}{c_s^2} \frac{d\phi}{dr} \ .
\label{flow_eqC3}
\end{equation}
We consider an isothermal cold flow with temperature $T=10^4$~K, and sound speed, $c_s = [{\rm k_B} T / (\mu {\rm m_H})]^{1/2}$, where $\mu=0.6$ is the mean molecular weight, $m_{\rm H}$ is the mass of the hydrogen atom.   Eq.~\ref{flow_eqC3} leads to the following family of solutions,
\begin{equation}
\frac{\mathcal{M}^2}{2} - \ln{\mathcal{M}} = 2 \ln{r} - \frac{\phi}{c_s^2} + C, 
\label{solutionC3} 
\end{equation}
where $C$ is an arbitrary constant.

\begin{figure}
    \centering
    \includegraphics[width=\linewidth]{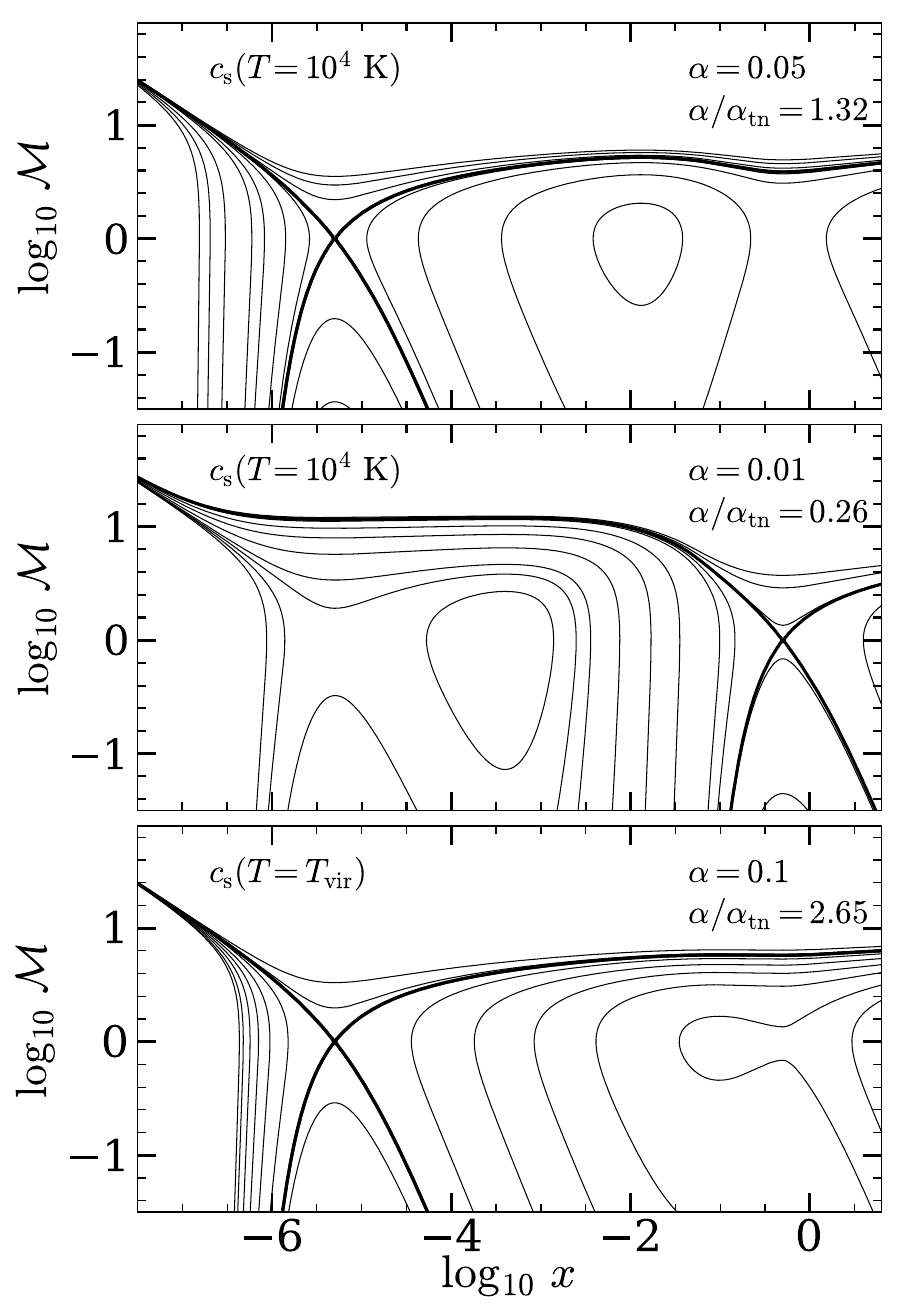}
    \caption{Families of solutions for Mach number, $\mathcal{M}$ as a function of dimensionless radial distance, $x= r c_{\rm s}^2/(G M_{\rm h})$ is shown as contours drawn by varying $C$ in Eq.~\ref{solutionC3}, for a fiducial value of $\beta=10^{-5}$, for $T=10^4$~K, $\alpha=0.05$ (top panel), for $T=10^4$~K, $\alpha=0.01$ (middle panel) and for $T=T_{\rm vir}$, $\alpha=0.1$ (bottom panel). Transonic solutions are highlighted by using the thick black lines. The corresponding $\alpha/\alpha_{\rm tn}$ values are also shown in each panel.}
    \label{fig_Mach_x}
\end{figure}

As shown in \cite{Sharma_2024}, this system admits three critical points, obtained by setting the right hand side of Eq.~\ref{flow_eqC3} to zero. The transonic solution (from Eq.~\ref{solutionC3}) passes through the innermost critical point when the black hole controls the accretion, and through the outermost critical point when the halo governs the accretion.  Two crucial parameters determine the relative effects of the black hole and the halo. First parameter is $\alpha = R_{\rm s} c_{\rm s}^2 /(G M_{\rm h})$, where $M_{\rm h}$ is the halo mass, $R_{\rm s} = R_{\rm vir}/c$ is the NFW scale radius; $R_{\rm vir}$ is the halo virial radius and $c$ is the concentration parameter. Thus, $\alpha= T/(2 c T_{\rm vir})$, where $T$ is flow temperature and $T_{\rm vir}= \mu m_{\rm H} G M_{\rm h}/(2 {\rm k_B} R_{\rm vir})$ is the halo virial temperature. The inverse of $\alpha$ gauges the strength of the halo gravity against the gas pressure, and its value depends on the gas temperature relative to the virial temperature of the halo. The second crucial parameter is $\beta = M_{\rm bh}/ M_{\rm h}$, which is the ratio of black hole mass to the halo mass.  

Fig.~\ref{fig_Mach_x} shows the family of solutions corresponding to Eq.~\ref{solutionC3} for $\alpha=0.05$ (top panel), $0.01$ (middle panel), and $0.1$ (bottom panel) for a fiducial value of $\beta=10^{-5}$. There are two distinct possibilities whereas the solution either passes through the innermost critical point for high $\alpha$ (top and bottom panels) or through the outermost critical point for low $\alpha$ (middle panel).   In the top panel, the gas temperature, $T=10^4$~K, with  $\alpha=0.05$, and the solution passes through the innermost critical point. As $\alpha$ is decreased to $0.01$ (middle panel) the halo gravity becomes dominant and the solution passes through the outermost critical point. However, in the bottom panel we vary the gas temperature and set it equal to the virial temperature, leading to a high $\alpha=0.1$, and again the flow transitions  and passes through the innermost critical point.
The accretion rate is given by, 
\begin{equation}
\dot M_{\rm bh} =  \frac{4 \pi G^2 \rho_{\infty} M_{\rm h}^2}{c_{\rm s}^3} \lambda
\label{eq.M_accnfwC3}
\end{equation}
where $\lambda$ is the dimensionless accretion rate given by, 
\begin{equation}
\lambda(\alpha,\beta) \approx \begin{cases}
\frac{\beta^2}{4} e^{1/(\alpha\eta)}  e^{3/2} \approx 1.12 \beta^2 e^{1/(\alpha\eta)} &\text{$\alpha > \alpha_{\rm tn}$}\\
\frac{(1+\beta)^2}{4} e^{3/2} \approx 1.12 &\text{$\alpha < \alpha_{\rm tn}$}
\end{cases}
\label{eq_lamb_nfw_C3}
\end{equation} 
The parameter $\eta = (1+1/c)\ln{(1+c)} - 1$, and $\alpha_{\rm tn}$ is the transition value of $\alpha$ that separates the two regimes, given by, $\alpha_{\rm tn} \approx [2 \eta \ln{\left(1/\beta \right)}]^{-1}$. The classic Bondi result $\lambda=1.12$ \citep{Bondi1952} is apparent in the two extreme limits. For high $\alpha \gg \alpha_{\rm tn}$, the black hole controls the accretion, and for low $\alpha \ll \alpha_{\rm tn}$, the halo gravity is dominant and the halo controls the accretion.

The parameter $\alpha$ depends on the halo mass as $\alpha \propto R_{\rm vir}/M_{\rm h}$. Considering roughly $R_{\rm vir} \propto M_{\rm h}^{1/3}$,  we get 
$\alpha \propto M_{\rm h}^{-2/3}$.  As the halo mass evolves with redshift, $\alpha$ evolves along with it.  Thus,  $\alpha$ is initially high as in the top panel of Fig.~\ref{fig_Mach_x}, but then it decreases with redshift as the halo mass increases (middle panel). Therefore, the accretion rate initially is controlled by the black hole. Thereafter, at a specific redshift, $\alpha$ approaches $\alpha_{\rm tn}$, and a transition to a high accretion rate due to the dark matter halo occurs, resulting in a several e-fold increase in the black hole mass over a short span of time \citep{Sharma_2024}. Similarly, as $\alpha \propto  T$, therefore a transition can also occur if the gas temperature evolves.

To calculate the evolution of black hole mass, we can integrate the accretion rate in cosmological settings where the dark matter halo mass evolves with  redshift \citep{Correa1}. The following integral provides the black hole mass as a function of redshift, 
\begin{equation}
    M_{\rm bh}(z) = \int_{z_{\rm h}}^{z}  \frac{-\dot M_{\rm bh}}{(1+z') H(z')} dz'
    \label{eqaccinteg}
\end{equation}
where $H(z)$ is the Hubble parameter. We consider a stellar seed of $10$~M$_\odot$ that begins to grow at an early redshift $z_{\rm h}=25$. We use the evolution of halo masses with redshift from \cite{Correa1, Correa-mnras}, which can  be described by $M_{\rm h}(z) = M_{\rm h,0} (1+z)^{a_{\rm h}(M_{\rm h,0})} e^{-b_{\rm h}(M_{\rm h,0})}$ \citep{Correa1}.  We take $\rho_{\infty} = 10^{-2}  \rho_{\rm b}$, where $\rho_{\rm b}$ is the baryon density in the Universe. We use \citet{Planck2015} cosmological parameters, and a typical value of concentration parameter $c = 5$ for high redshift haloes \citep{Prada12, Correa-mnras}.


\subsection{Origin of OBHs and LRDs}
\begin{figure}[]
    \centering
    \includegraphics[width=\linewidth]{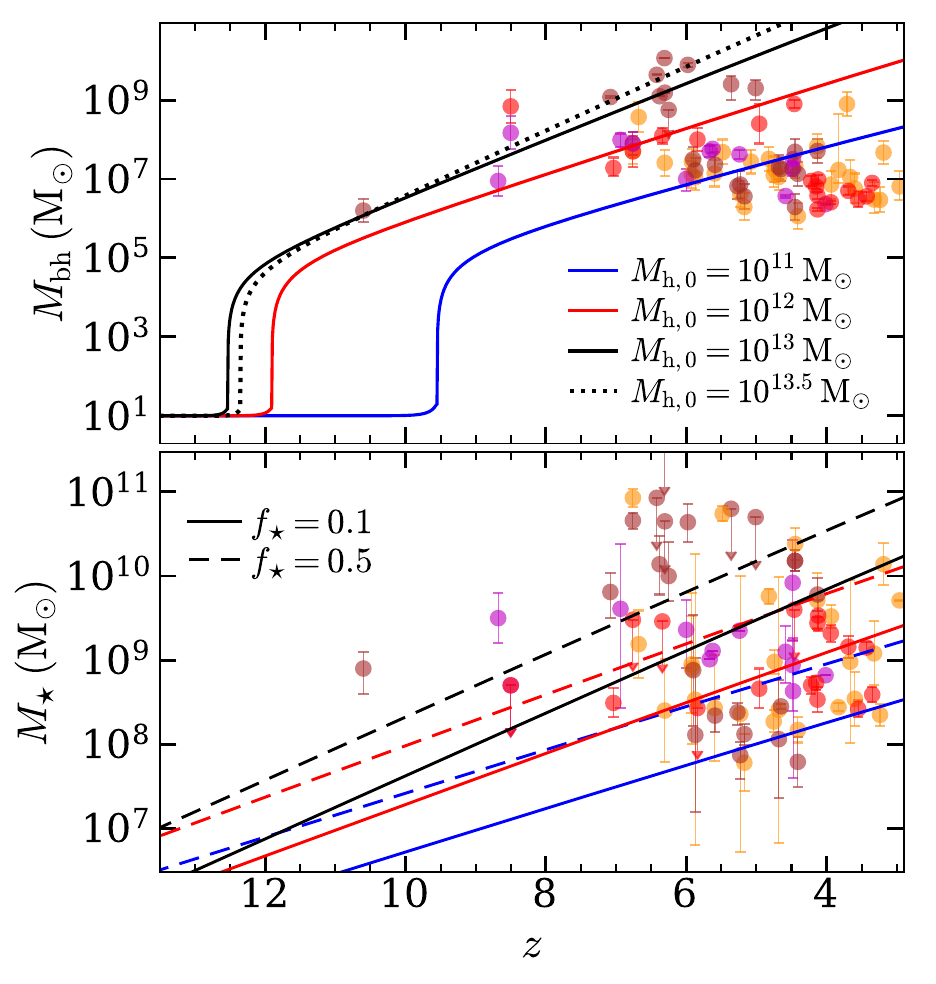}
    \caption{{\it Top Panel}: The redshift evolution of the central black hole mass, $M_{\rm bh}$, within dark matter haloes with present day mass $M_{\rm h, 0}= 10^{11}$ (solid blue curve), $10^{12}$ (solid red curve), $10^{13}$ (solid black curve) and $10^{13.5}$~M$_\odot$ (dotted black curve). The black holes observed by the JWST at $z\gtrsim 3$ are shown as magenta \citep{Harikane2023, Kokorev2023, Larson2023}, brown \citep{Maiolino2024a, Maiolino2024b, Stone2024, Yue2024}, orange \citep{Juodzbalis2026} filled circles. The LRDs that are increasingly interpreted as AGN are shown as red circles \citep{Akins2025, Chang2025, Pang2026}. {\it Bottom Panel}:  Evolution of the stellar masses corresponding to the black hole masses in the top panel. The stellar masses are obtained by using  $M_{\star}= f_\star f_{\rm b} M_{\rm h}$ where $f_{\rm b}$ is the baryon fraction \citep{Planck2015}, and  $f_\star$ is the fraction of baryons converted to stars. The solid curves are for $f_\star=0.1$ and dashed curves for $f_\star=0.5$. The JWST observations (filled circles) with stellar masses corresponding to the black hole masses in the top panel are also shown for comparison.} 
    \label{fig.Bh-z_const}
\end{figure}

In the top panel of Fig.~\ref{fig.Bh-z_const}, we show the evolution of $M_{\rm bh}$ for   haloes with present day masses $M_{\rm h,0} = 10^{11}$, $10^{12}$, $10^{13}$ and $10^{13.5}$~M$_\odot$  shown as solid blue,  solid red, solid black and dotted black curves, respectively. The filled circles represent OBHs \citep{Harikane2023, Maiolino2024a, Maiolino2024b, Stone2024, Yue2024,  Juodzbalis2026} and LRDs that are AGN candidates \citep{Akins2025, Chang2025, Pang2026} detected recently by JWST at $z\sim 3-7$. At a specific redshift which depends on the halo mass, the accretion transitions to the halo-driven rapid phase leading to massive black holes that explains the observations.

In high-mass haloes, the transition occurs early, while in low-mass haloes it occurs late. For $M_{\rm h,0} = 10^{13}$~M$_\odot$, the transition occurs at $z\approx 12$, for $M_{\rm h,0} = 10^{11}$~M$_\odot$ at $z\approx 9$, while for $M_{\rm h,0} = 10^{10}$~M$_\odot$ it occurs at $z<6$. During rapid growth phase, the accretion is near- or super-Eddington \citep{Sharma_2024} which is consistent with observed Eddington ratios for high redshift SMBHs \citep{Onoue2019, Fan2023}.

The growth histories show diminishing differences as $M_{\rm h,0}$ increases, following the adopted halo mass histories 
\citep{Correa1, Correa-mnras}, in which the masses of massive haloes at a given high redshift vary weakly with $M_{\rm h,0}$. At $z\gtrsim 10$ the histories of the $M_{\rm h,0}=10^{13}$ and 
$10^{13.5}$~M$_\odot$ haloes cross, reflecting the uncertainty of halo mass histories at the highest masses and redshifts.

\begin{figure}[]
    \centering
    \includegraphics[width=\linewidth]{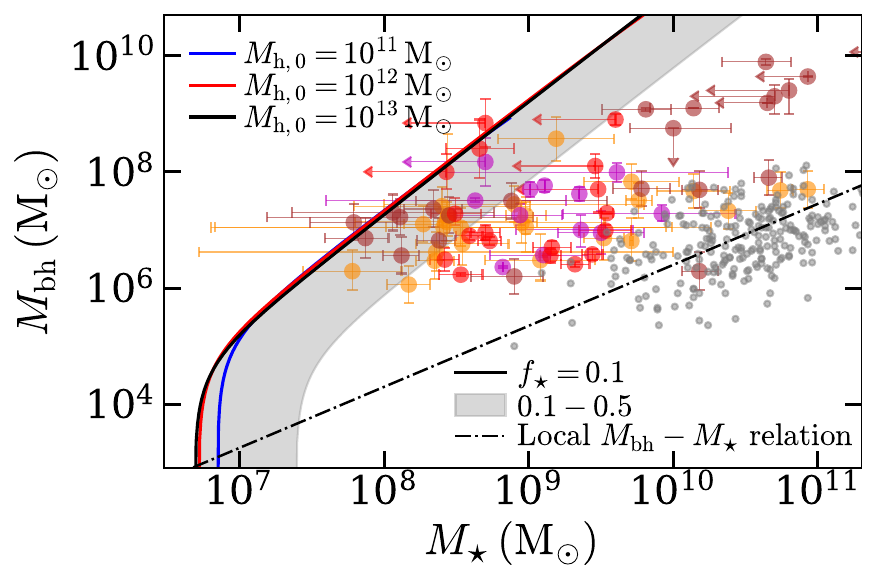}
    \caption{Evolution of the black hole mass, $M_{\rm bh}$, with the stellar mass, $M_{\star}$, in the halo-driven growth model, for haloes with present day masses, $M_{\rm h, 0}=10^{11}$ (blue curve), $10^{12}$ (red curve) and $10^{13}$ M$_\odot$ (black curve). The solid lines correspond to $f_\star=0.1$, and the shaded region shows the range $f_\star=0.1\hbox{--}0.5$. The local $M_{\rm bh}\hbox{--}M_\star$ relation is shown as grey dots and a black dash-dotted line \citep{Reines2015}. The recent JWST observations are shown as magenta \citep{Harikane2023, Kokorev2023, Larson2023}, brown \citep{Maiolino2024a, Maiolino2024b, Stone2024, Yue2024} and orange circles \citep{Juodzbalis2026}, and the red solid circles represent the Little Red Dots \citep{Akins2025, Chang2025, Pang2026}.}
    \label{fig.Mbh-Mstar_main}
\end{figure}

The stellar masses corresponding to the JWST black holes shown in Fig.~\ref{fig.Bh-z_const}, top panel, are also available. To compare, we can calculate the stellar mass theoretically for our model using $M_{\star}= f_\star f_{\rm b} M_{\rm h}$, where $f_{\rm b}$ is the baryon fraction in the Universe. $f_\star$ is the cumulative star formation efficiency, which is the fraction of baryons converted to stars. The star formation in early galaxies under surveillance by JWST is typically bursty \citep{Robertson2024, Looser2025, Joshi2025}, and the star formation efficiency likely fluctuates and can be unusually high \citep{Dekel2023, Chen2026} compared to expectations from 
traditional low-redshift studies \citep{Behroozi2013, Moster2013, Lacey2016}. In  Fig.~\ref{fig.Bh-z_const} bottom panel, we show the evolution of stellar masses in dark matter haloes for a conservative $f_\star=0.1$ (solid lines), and its variation to higher values, $f_\star=0.5$ (dashed lines), and validate with the JWST observations (colour filled circles).

The OBHs detected by JWST \citep{Maiolino2024a, Maiolino2024b, Stone2024, Yue2024} and the OBHs among LRDs \citep{Chang2025, Pang2026}, typically have masses higher than expected from the local black hole-stellar mass relation \citep{Reines2015, Harikane2023}. In Fig.~\ref{fig.Mbh-Mstar_main}, for our theoretical model, we show 
the $M_{\rm bh}$--$M_\star$ relation using solid lines for $f_\star=0.1$. We compare our results with the JWST observations as well as with the local stellar mass--black hole mass correlation. The fraction $f_\star$ is known to depend on the halo mass 
\citep{Behroozi2013, Moster2013, Behroozi2019}, and it also likely evolves with redshift \citep{Dekel2023, Harikane2023galaxy, Mason2023, Shen2026}. The dependence of $f_\star$ on halo mass and redshift has been studied extensively using semi-analytical models and empirical constraints \citep{Behroozi2019, Sharma2020, Zhang2023, Scoggins2024, Wang2025, Hu2026}. To account for possible variations due to the evolution of $f_\star$, we consider a range of $f_\star = 0.1$--$0.5$, shown as the gray shaded region in Fig.~\ref{fig.Mbh-Mstar_main}. Varying $f_\star$ covers the scatter in observational data-points well.

We recover a steeper relation between black hole mass and stellar mass, and our model explains the OBHs at high redshifts. If the black hole growth in our model continues, the masses become unusually high (the black curves in Fig.~\ref{fig.Bh-z_const} and \ref{fig.Mbh-Mstar_main}). However, eventually, the growth of black holes is expected to be regulated as we explore in the next section.

\section{Transition to Halo-thermodynamics driven suppression}
\label{sec_suprresion}
The rapid early growth of black holes must eventually be regulated to avoid unlimited mass accumulation leading to unphysical black hole masses. A variety of quenching mechanisms have been proposed, including radiative and mechanical feedback from accreting matter, and merger-driven star formation and associated gas depletion \citep{Rees_Ostriker1977, Silk1977, Silk98, Proga2000, King03, DIMatteo2005,  Hopkins2011, Fabian2012}.

We consider a fundamental mechanism governed by halo thermodynamics, based on the well-established interplay between cooling and dynamical time-scales that leads to the formation of stable virial shocks in haloes above a critical mass \citep{Birnboim03, Dekel2006, Keres2005, Cattaneo2006, Croton2006, Bower17}. In this regime, gas is shock-heated to the virial temperature and the supply of cold, rapidly accreting material to the central regions is significantly reduced, thereby suppressing black hole growth.

The evolution in the halo thermodynamics is an outcome of the interplay between the free-fall or the dynamical time, and the cooling time. The free-fall time of gas in haloes is typically, $t_{\rm ff} \sim (G\rho_{\rm h})^{-1/2}$, where $\rho_{\rm h} \sim 200\rho_{\rm crit}$ is the characteristic virial density and $\rho_{\rm crit}$ is the critical density of the Universe. The cooling time is roughly, $t_{\rm cool} \sim k_{\rm B} T / (n\Lambda(T))$, where $\Lambda(T)$ is the equilibrium cooling function  \citep{Sutherland1993} and $n$ is the baryon density. The ratio $t_{\rm cool}/t_{\rm ff}$ dictates whether the halo gas is cold or hot. For $t_{\rm cool}\ll t_{\rm ff}$, cooling is efficient and the gas settles to the atomic cooling floor $T\sim 10^4$~K. For $t_{\rm cool}\gtrsim t_{\rm ff}$, cooling becomes ineffective, the gas is shock-heated to the virial temperature $T\sim T_{\rm vir}$, and the halo enters the hot mode.
We can use a sigmoid function to approximate this behaviour to describe the temperature of the gas at the periphery of the halo as,
\begin{equation}
    T=10^4~{\rm K} + T_{\rm vir}\tanh{[(t_{\rm cool}/t_{\rm ff})^k]} \ ,
    \label{eq_tratio}
\end{equation} where $
t_{\rm cool}/t_{\rm ff} \sim k_{\rm B} T (G \rho_{\rm h})^{1/2}/[{n\Lambda(T)}] $ is evaluated at $T=T_{\rm vir}$, and $k$ is a parameter that controls the sharpness of transition. 
In the regime $T=10^5\hbox{--}10^{6.5}$~K, the cooling function can be approximated as, $\Lambda(T) \propto T^{-0.5}$ \citep{Sutherland1993}. In the matter dominated case at high redshift, both $\rho_{\rm h}$ and $n$ roughly scale with redshift as  $\propto (1+z)^3$.  Therefore, 
$
t_{\rm cool}/t_{\rm ff} \propto T_{\rm vir}^{1.5}(1+z)^{-1.5} \propto M_{\rm h}(z)
$. 
Thus, the criterion, $t_{\rm cool} \ll t_{\rm ff}$, for rapid cooling translates to an equivalent criterion on halo mass, which is $M_{\rm h}\ll M_{\rm h,c}$, where $M_{\rm h,c}$ is the critical halo mass at which the cooling is ineffective and the halo switches to the hot mode \citep{Rees_Ostriker1977, Silk1977,  WhiteRees1978, White1991, Birnboim03, Keres2005, Dekel2006}. We therefore implement the thermodynamic behaviour of the halo gas in our models as 
\begin{equation}
    T= 10^4~ {\rm K} + T_{\rm vir} \tanh{[(M_{\rm h}/M_{\rm h,c})^k]} \ , 
    \label{eq_mcratio}
\end{equation}
where $M_{\rm h,c}$ is defined by $t_{\rm cool}=t_{\rm ff}$, so that 
$t_{\rm cool}/t_{\rm ff}=M_{\rm h}/M_{\rm h,c}$. This impacts the sound 
speed $c_{\rm s}$ and $\alpha$ (Eq.~\ref{eq.M_accnfwC3} and 
\ref{eq_lamb_nfw_C3}), leading to the suppressed growth of black holes 
as the haloes switch to the hot mode.

The parameter $M_{\rm h,c}$ denotes the critical halo mass above which stable virial shocks develop, marking the transition from cold to hot mode and leading to a reduction in the cold gas supply to the centre \citep{Birnboim03}. \citet{Dekel2006} further investigated metallicity-dependent cooling and showed that virial shocks develop in haloes as low as $M_{\rm h, c} \sim 10^{11}-10^{12}$~M$_\odot$. In the present-day Universe, $M_{\rm h,c}\approx 10^{12}$~M$_\odot$, although theoretical arguments and simulations suggest that $M_{\rm h,c}$ decreases with increasing redshift.  In low-metallicity gas at high redshift,  stable virial shocks may form in haloes with masses as low as $\sim 10^{11}$~M$_\odot$ \citep{Dekel2006}, and transition to the hot mode may occur at such low halo masses \citep{Keres2005}. Motivated by these results, we adopt $M_{\rm h,c}$ in the range $10^{11}$--$10^{12}\,\mathrm{M}_\odot$ in our models.

The parameter $k$ describes how steeply the temperature rises to 
the virial temperature, or how rapidly the cooling of the halo gas 
becomes ineffective. A high value of $k$ represents a sharp transition 
at $t_{\rm cool} \simeq t_{\rm ff}$, corresponding to the cooling 
criterion of \citet{Rees_Ostriker1977} and \citet{Silk1977}. However, 
the actual transition proceeds through an interplay of shock heating, 
virialization and cooling, and the detailed studies of \citet{Dekel2006} (see also \citealt{Birnboim03, Keres2005}) indicate that it is gradual. We therefore adopt $k=1$ in our main models, motivated by these studies, and also explore the effect of varying $k$ in the next section.

\section{Results}
\label{sec_result}

\begin{figure}
    \centering
    \includegraphics[width=\linewidth]{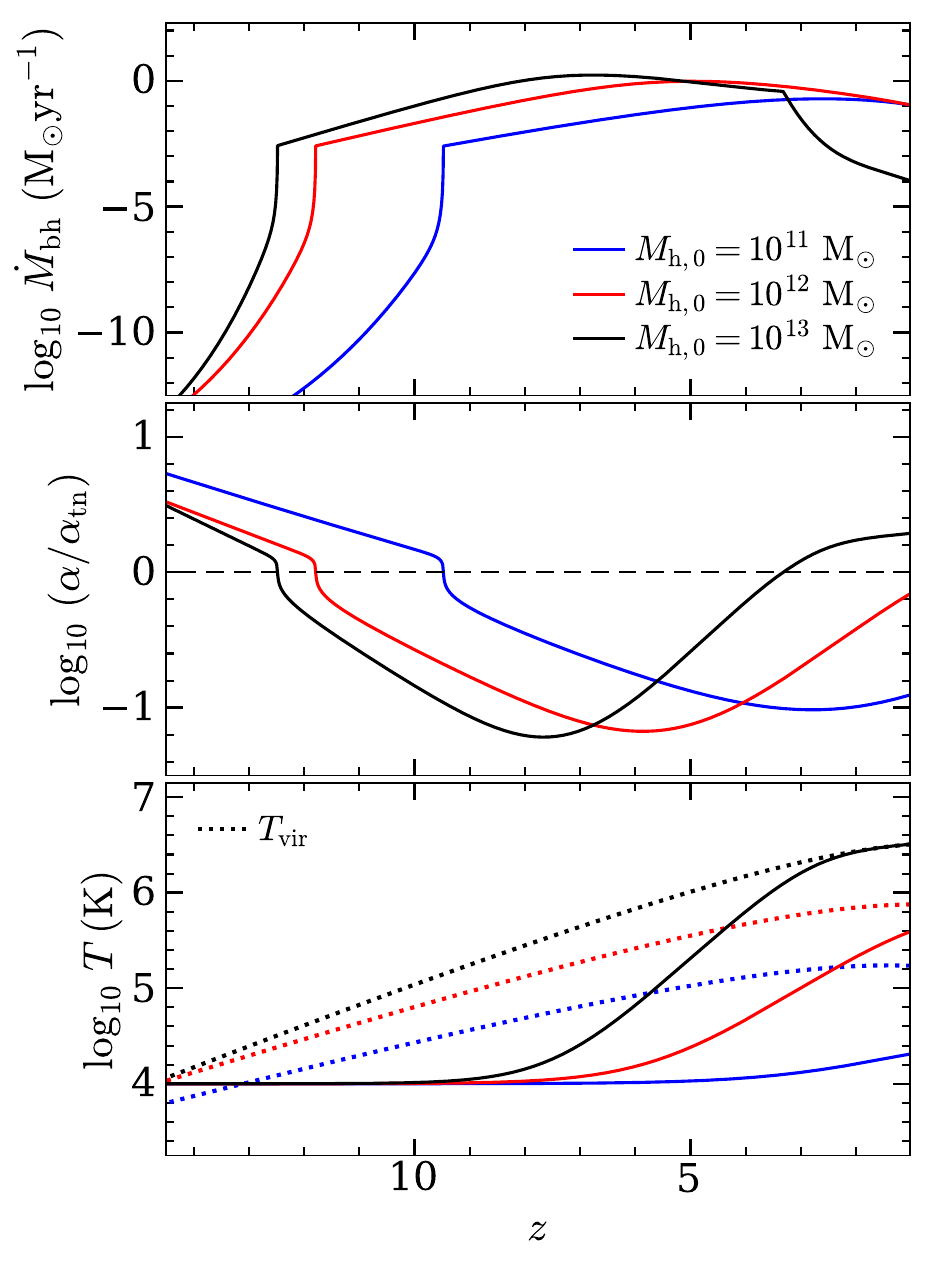}
    \caption{{\it Top panel}: Evolution of the mass accretion rate, $\dot{M}_{\rm bh}$ with redshift, $z$, for halo-driven growth followed by suppression due to inefficient cooling and shock heating of the halo gas as the halo mass approaches $M_{\rm h, c}= 10^{12}$~M$_{\odot}$, is shown for haloes with present day mass $M_{\rm h, 0}= 10^{11}$ (blue curve), $10^{12}$ (red curve), and $10^{13}$~M$_{\odot}$ (black curve). {\it Middle panel}: Redshift evolution of the ratio $\alpha/\alpha_{\rm tn}$, where $\alpha$ is the halo parameter and $\alpha_{\rm tn} = [2\eta \ln{(1/\beta)}]^{-1}$ is its transition value that demarcates the transition between rapid (for $\alpha<\alpha_{\rm tn}$)  and slow (for $\alpha>\alpha_{\rm tn}$) black hole growth. The condition, $\alpha=\alpha_{\rm tn}$, is shown as a horizontal dashed line. {\it Bottom Panel}: Evolution of the flow temperature $T$ for the same three haloes as in the upper two panels. The corresponding virial temperature  evolution is shown as dotted lines for comparison.}    
    \label{fig_MaccT}
\end{figure}

We integrate the accretion rate (Eq.~\ref{eq.M_accnfwC3} and 
\ref{eqaccinteg}) to follow the growth of black holes in haloes evolving with redshift. In Fig.~\ref{fig_MaccT}, we show the evolution of the black hole accretion rate (upper panel) and the evolution of the ratio $\alpha/\alpha_{\rm tn}$ (middle panel) where $\alpha$ is the halo parameter and $\alpha_{\rm tn}$ is its critical transition value.  The condition $\alpha=\alpha_{\rm tn}$ is shown as a horizontal dashed line. In the cold mode, $T=10^4$~K and the sound speed is constant, therefore $\alpha \propto M_{\rm h}^{-2/3}$. As the halo grows with time, $\alpha$ decreases and eventually falls below the transition value $\alpha_{\rm tn} \approx [2\eta\ln(1/\beta)]^{-1}$, triggering a hydrodynamic switch from the innermost to the outermost critical point driven by halo gravity. The accretion rate thus switches rapidly to high values in the halo-gravity dominated phase, giving rise to supermassive black holes. We further see that $\alpha$ decreases to a minimum but then turns around and begins increasing as the halo mass approaches $M_{\rm h,c}$ and the halo transitions to the hot mode ($T$ approaches $T_{\rm vir}$) leading to suppression of black hole growth. For the most massive haloes (black curve), $\alpha$ rises back above $\alpha_{\rm tn}$.

The suppression occurs in two successive stages within the same hydrodynamic framework. As $\alpha$ passes its minimum, suppression begins as the halo 
transitions to the hot mode and the temperature approaches $T_{\rm vir}$. As a result, accretion rates plateau and gently decrease (upper panel, Fig.~\ref{fig_MaccT}), consistent with the dependence of the accretion rate on sound speed in Eq.~\ref{eq.M_accnfwC3}. At $z\approx 3$, the accretion rate for $M_{\rm h,0}=10^{13}$~M$_{\odot}$ (black curve) drops abruptly as $\alpha$ rises above $\alpha_{\rm tn}$, triggering a second hydrodynamic transition back to the innermost critical point (Eq.~\ref{eq_lamb_nfw_C3}), which vigorously suppresses black hole growth further. 

Thus, the black hole growth features a transient phase of rapid growth 
sandwiched between two key transitions. First, a halo-gravity-assisted 
transition leads to rapid growth and the origin of overmassive black 
holes; subsequently, a halo-thermodynamics-driven transition begins with a gradual decline in accretion rate as the halo gas fails to cool and approaches the virial temperature, followed by vigorous suppression once $\alpha\gtrsim\alpha_{\rm tn}$. During this transient phase, the high accretion rate produces black holes that are overmassive relative to the prevailing local scaling relations.

\begin{figure}
    \centering
    \includegraphics[width=\linewidth]{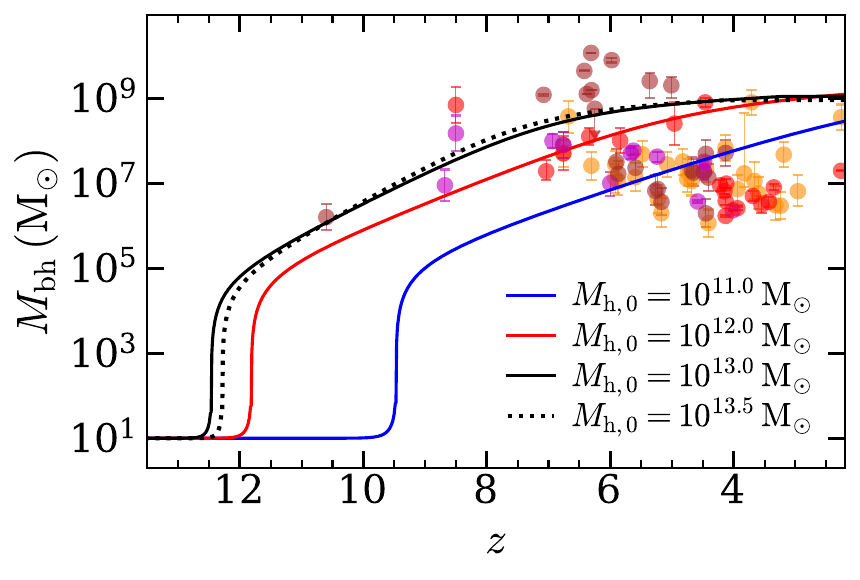}
    \caption{Evolution of the central black hole mass $M_{\rm bh}$ with redshift, $z$, in the halo driven growth model with suppression due to virial shocks for dark matter haloes having present day masses $M_{\rm h, 0}= 10^{11}$, $10^{12}$, $10^{13}$ and $10^{13.5}$~M$_\odot$ is shown as solid blue, solid red, solid black and dotted black curves, respectively. The critical halo mass for suppression is $10^{12}$~M$_\odot$. The JWST observations at $z\gtrsim 3$ are shown as magenta \citep{Harikane2023, Kokorev2023, Larson2023}, brown \citep{Maiolino2024a, Maiolino2024b, Stone2024, Yue2024}, orange filled circles \citep{Juodzbalis2026}. The red filled circles represent the Little Red Dots \citep{Akins2025, Chang2025, Pang2026}.
    }
    \label{fig_crit_z}
\end{figure}

\begin{figure}
    \centering
     \includegraphics[width=\linewidth]{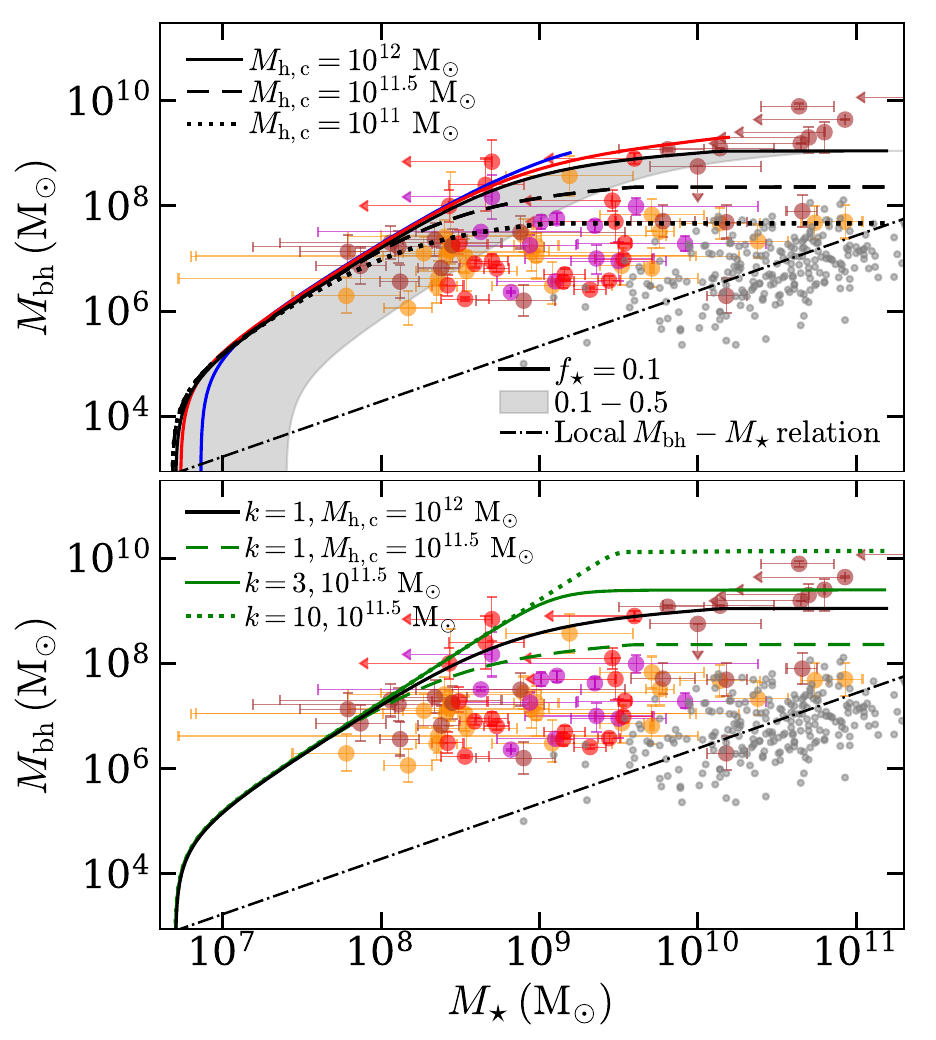}
    \caption{{\it Top panel}: Evolution of the black hole mass, $M_{\rm bh}$, with the stellar mass, $M_{\star}$, from the halo driven growth model with suppression at a critical value of halo mass, $M_{\rm h,c}=10^{12}$~M$_\odot$, is shown for  three haloes with present day masses, $M_{\rm h, 0}=10^{11}$ (solid blue curve), $10^{12}$ (solid red curve) and $10^{13}$ M$_\odot$ (solid black curve). The stellar mass is obtained by using, $M_{\star}= f_\star f_{\rm b} M_{\rm h}$, where $f_\star$ is the fraction of baryons converted to stars and $f_{\rm b}$ is the cosmic baryon fraction \citep{Planck2015}. For $M_{\rm h,0}= 10^{13}$ M$_\odot$, we have shown the effect of varying $M_{\rm h, c}$, from $10^{12}$~M$_\odot$ (solid black curve) to $10^{11.5}$ (dashed black curve) and $10^{11}$ M$_\odot$ (dotted black curve). Moreover, the solid curves correspond to $f_\star=0.1$, and the shaded region shows the range $0.1-0.5$. The local $M_{\rm bh}\hbox{--}M_\star$ relation is shown with grey dots and a black dash-dotted line \citep{Reines2015}. The JWST data-points are shown as magenta \citep{Harikane2023, Kokorev2023, Larson2023}, brown \citep{Maiolino2024a, Maiolino2024b, Stone2024, Yue2024}, orange filled circles \citep{Juodzbalis2026}, and the Little Red Dots are shown as the red filled circles \citep{Akins2025, Chang2025, Pang2026}. {\it Bottom panel}: same as the top panel but for variations in our suppression scheme due to the shock heating of the halo (Eq. \ref{eq_mcratio}) with $k=1$ (dashed green), $3$ (solid green), and $10$ (dotted green) for $M_{\rm h,c}=10^{11.5}$~M$_\odot$. The default $k=1$ model from the top panel with $M_{\rm h,c}=10^{12}$~M$_\odot$ is also shown as a solid black curve. 
    }
    \label{fig_crit_mstar}
\end{figure}

In Fig.~\ref{fig_crit_z}, we show the calculated evolution of 
$M_{\rm bh}$, which compares  well with the JWST data. There is a decrease in the black hole mass, specifically visible for the massive haloes (black curves), in comparison to Fig.~\ref{fig.Bh-z_const}, which is due to the suppression in growth as the halo enters the hot mode. Some of the observed systems lie above the solid black curve ($M_{\rm h,0}=10^{13}$~M$_\odot$). The most extreme observed systems likely reside in the rarest, most massive environments. Furthermore, black hole mass estimates at $z\gtrsim 5$ carry substantial 
observational uncertainties of $0.3$--$0.5$~dex, which must be considered when comparing individual data points to theoretical curves. Our theoretical black curve approaches $10^9$~M$_\odot$, over the redshift range $z\approx 5$--$6$.

The black hole growth trajectories also approach an effective upper 
limit regardless of the present day halo mass. In the halo-dominated 
regime, $\dot{M}_{\rm bh} \propto M_{\rm h}^2/c_{\rm s}^3$ 
(Eq.~\ref{eq_lamb_nfw_C3}), which favours more massive haloes. 
However, the suppression begins as the halo reaches the critical mass 
$M_{\rm h,c}$, irrespective of its present day mass, 
and the more massive haloes enter 
the hot mode earlier, at higher redshifts. The competition between these two effects produces a ceiling on black hole growth that is set largely by $M_{\rm h,c}$, at $M_{\rm bh} \sim 10^{9}$~M$_\odot$ for $M_{\rm h,c}=10^{12}$~M$_\odot$.

In Fig.~\ref{fig_crit_mstar} (top panel), we show the black hole to stellar mass relation for our model (curves and the shaded region), which explains the unusually high masses of JWST-detected OBHs and LRDs. Further, our model exhibits a suppression causing the black holes growth to settle towards the local coevolutionary trend.  

Thus, OBHs and LRDs emerge due to rapid early growth of black holes triggered  by host haloes.  The rapid phase of growth begins at a redshift that depends on  the halo mass. For example, the growth transitions to rapid rates at about $z\gtrsim 12$ for $M_{\rm h,0}\ge 10^{12}$~M$_\odot$. Interestingly, the rapid growth begins when the stellar mass is roughly $10^7$~M$_\odot$ for $f_\star=0.1$, irrespective of the halo mass. 

After the haloes pass through the transient OBH or LRD phase, the halo-thermodynamics leads them into the `hot mode' dominated by formation of stable virial shocks \citep{Dekel2006}, which causes a suppression of the black hole growth. The combination of rapid growth and suppression explains the formation of OBHs and their subsequent evolution to the local $M_{\rm bh}-M_\star$ relation.

In Fig.~\ref{fig_crit_mstar} (bottom panel) we show the effect of 
varying the sharpness of the transition to heating, from a shallow 
transition \citep{Dekel2006} with $k=1$ to a sharp transition when $t_{\rm cool}\sim t_{\rm ff}$ as in \citet{Rees_Ostriker1977, Silk1977} with $k=10$ 
(Eq.~\ref{eq_tratio} and \ref{eq_mcratio}). For a suitable combination 
of $k$ and $M_{\rm h,c}$ the model gives a similar outcome for the JWST data. Our main model, with $k=1$ and $M_{\rm h,c}=10^{12}$~M$_\odot$, is motivated by the shock heating and virialization of cosmological haloes 
\citep{Birnboim03, Keres2005, Dekel2006}.

\begin{figure}[]
    \centering
    \includegraphics[width=\linewidth]{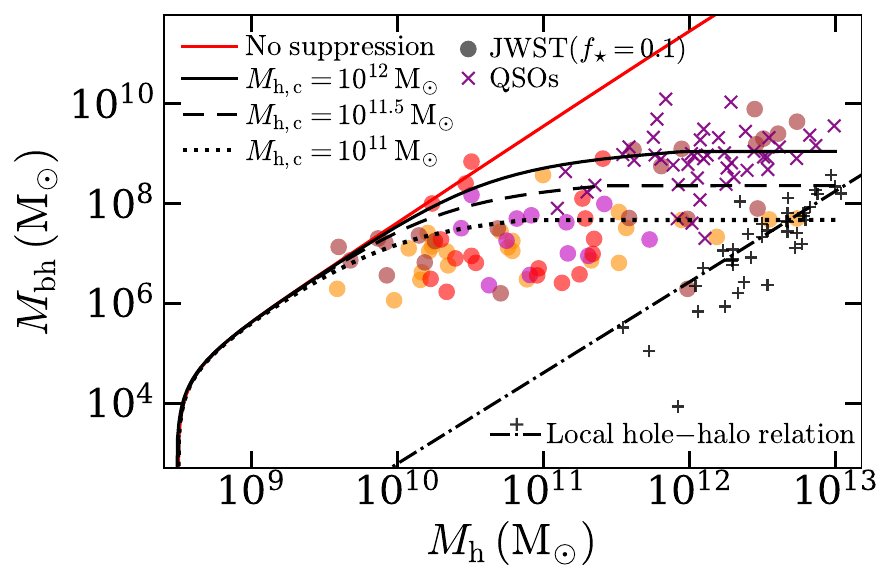}
    \caption{Evolution of the black hole mass, $M_{\rm bh}$ with halo mass, $M_{\rm h}$ in the halo-driven growth model, for a halo with present-day mass $M_{\rm h, 0}=10^{13}$ M$_\odot$, is shown as a solid red curve. The models with both growth and suppression with $M_{\rm h, c}= 10^{11}$ (dotted black curve), $10^{11.5}$ (dashed black curve) and $10^{12}$ M$_\odot$ (solid black curve), where $M_{\rm h,c}$ is the critical value of halo mass at which the suppression ensues. The correlation observed in the local Universe is shown as a black dash-dotted line and black plus symbols \citep{Ferrarese02}. The quasars from \cite{Shimasaku2019} at $z\sim 6$ are shown as purple cross symbols. To compare with the recent JWST observations, we infer the halo mass corresponding to the reported stellar mass by using $M_{\rm h} = M_\star/(f_\star f_{\rm b})$ with $f_\star=0.1$. The JWST SMBHs are shown as magenta \citep{Harikane2023, Kokorev2023, Larson2023}, brown \citep{Maiolino2024a, Maiolino2024b, Stone2024, Yue2024} and orange filled circles \citep{Juodzbalis2026}. The LRDs are shown as red filled circles \citep{Akins2025, Chang2025, Pang2026}. }
    \label{fig.Mbh-Mh_ferrarese}
\end{figure}

Our theoretical model also explains the coevolution of black hole masses with their host haloes shown as red and black curves in Fig.~\ref{fig.Mbh-Mh_ferrarese}, compared with the observed high-redshift quasars with known halo masses \citep{Shimasaku2019} (purple crosses), and also with the latest JWST detected OBHs and LRDs for which we used a fiducial conversion factor $f_\star=0.1$ to obtain representative halo masses from the measured stellar masses (solid circles). The halo-driven rapid growth models overproduce black hole masses relative to the local hole-halo correlation \citep{Ferrarese02}, shown as a black dash-dotted line in agreement with high-redshift quasars and JWST observations. Interestingly, the rapid growth model without suppression (red curve) has a slope $\approx 2$ as expected from Eq.~\ref{eq.M_accnfwC3}, which is roughly the same as the local relation, while the slopes of the models with suppression  decrease and the system evolves toward the local correlation.

\section{Discussion and Conclusions}
\label{sec_discussionC3}
Recent observations from JWST have revealed a population of high-redshift OBHs \citep{Harikane2023, Maiolino2024a, Maiolino2024b, Stone2024, Juodzbalis2026, Yue2024}, and LRDs that are increasingly interpreted as compact, gas-enshrouded accreting black holes \citep{Matthee2024, Chang2025, Pang2026, Rusakov2026, Hviding2026}. While LRDs represent a phenomenologically distinct population, both OBHs and a significant number of LRDs with known black hole masses lie significantly above the canonical local scaling relations \citep{Harikane2023, Pacucci2024, Li2025, Juodzbalis2026}.

To address the JWST observations, we have investigated hydrodynamic accretion within a combined black hole and NFW halo potential, finding that halo gravity drives an early rapid phase of central black hole growth (Fig.~\ref{fig.Bh-z_const}, \ref{fig.Mbh-Mstar_main}, \ref{fig_crit_z}, and \ref{fig_crit_mstar}). In this phase, gas at the atomic cooling limit, $T=10^4$~K, efficiently reaches the centre, allowing black holes to attain the high masses reported by JWST within the desired cosmological timescales.

Our results further indicate that a transition occurs once the host 
halo reaches a critical mass scale, $M_{\rm h,c}$, beyond which halo 
thermodynamics limits the gas supply. The onset of stable virial shocks 
\citep{Dekel2006} results in hot accretion flows and rapidly reduces 
black hole accretion. Furthermore, the suppression occurs in two 
successive stages. It begins as the sound speed rises toward 
$c_{\rm s}(T_{\rm vir})$, and the accretion rate drops as 
$\dot{M}_{\rm bh}\propto c_{\rm s}^{-3}$, followed by an abrupt drop to 
very low values as $\alpha$ rises back above $\alpha_{\rm tn}$, 
switching the transonic solution from the outermost (halo dominated) 
critical point back to the innermost (black hole dominated) critical 
point where $\lambda\sim\beta^2$.

The halo-scale inflow through cold streams may persist to some degree despite shock heating \citep{Keres2005, Dekel2006, vandeVoort2011, Paste2026}, in which case the suppression above $M_{\rm h,c}$ would be gentler than in our model.  Recent studies also indicate that the cold streams may themselves get heated and disrupted in the hot halo \citep{Hong2024, Medlock2026, Yao26}. For the adopted prescription with a constant star formation efficiency, the rapid phase of accretion followed by a smooth halo-mass-dependent suppression broadly reproduces the observed distribution of high-redshift systems in the $M_{\rm bh}$--$M_\star$ plane (Fig.~\ref{fig_crit_mstar}). Within this framework, the qualitative evolutionary picture is governed by halo gravity and halo thermodynamics.

The evolutionary trajectories predicted by our model pass through the OBH population during the early halo-driven rapid growth phase and subsequently bend toward the scaling relations observed in the local Universe \citep{Ferrarese02, Reines2015}. In this framework, LRDs may be an observational manifestation of the same rapid growth phase, although the two populations are not necessarily identical.

The present model adopts spherical symmetry, which is a standard simplifying approximation appropriate at halo scales where the large-scale potential is approximately isotropic. Similarly, placing a stellar-mass seed at the centre of the halo potential is a standard idealisation in accretion models \citep{Inayoshi2020}. The growth scenario presented in this study requires the seed black hole to reside at, or sink to, the centre of the halo potential before the transition to the hot mode. The settling timescale is governed by dynamical friction, and if this timescale exceeds the duration of the cold mode phase, the seed may not participate in the rapid growth episode.

In this work, the halo-scale gas reservoir feeds both star formation and black hole growth. In practice, star formation, angular momentum, and stellar feedback may reduce the fraction of halo-scale inflow that reaches the black hole, while AGN feedback in turn affects star formation.  The accretion rate in Eq.~\ref{eq.M_accnfwC3} and \ref{eq_lamb_nfw_C3} therefore represents an upper limit to black hole growth within the adopted model. The proposed difference between the responses of black hole growth and star formation to the hot mode transition is an assumption of the model and is not calculated self-consistently. In particular, the adopted prescription $M_\star = f_\star f_{\rm b} M_{\rm h}$ allows stellar mass to continue tracking halo mass through the hot mode transition, which contributes to the evolution toward the local $M_{\rm bh}$--$M_\star$ relation. A self-consistent treatment of these aspects is a natural extension of the present analytical framework, and cosmological simulations may explore them in detail.

Overall, our results highlight the central role of dark matter haloes in governing the early growth and subsequent regulation of SMBHs. The initial phase of cold-mode, halo-gravity-driven accretion produces the overmassive black holes observed at high redshift, while the transition to a hot, shock-supported halo leads to a decline in accretion and guides systems toward the coevolutionary relations seen in the local Universe.

\section*{Acknowledgements}
We thank an anonymous referee for constructive comments. RS thanks the ministry of education (MoE), Government of India, for support through a research fellowship. The authors gratefully acknowledge the financial support for open access publication provided by the One Nation One Subscription (ONOS) initiative, Department of Science and Technology (DST), Government of India.

\bibliography{APJ}{}
\bibliographystyle{aasjournalv7}

\end{document}